\documentclass{llncs}
\usepackage{makeidx}  % allows for indexgeneration
\usepackage{graphicx}

\usepackage{siunitx}
\usepackage{amsmath,amsfonts,amssymb}
\usepackage{caption}
\usepackage{subcaption}
\usepackage{color}
\usepackage{arydshln}

\usepackage{caption}
\makeatletter
\newcommand{\printfnsymbol}[1]{%
  \textsuperscript{\@fnsymbol{#1}}%
}
\makeatother

\begin{document}
\sloppy
\pagestyle{headings}  % switches on printing of running heads
\title{Automated Estimation of the Spinal Curvature via Spine Centerline Extraction with Ensembles of Cascaded Neural Networks}

\author{Florian Dubost\inst{1} \and
Benjamin Collery\inst{1,2}\thanks{equal contribution} \and
Antonin Renaudier\inst{1,2}\printfnsymbol{1} \and
Axel Roc\inst{1,2}\printfnsymbol{1} \and
Nicolas Posocco\inst{1,3} \and
Gerda Bortsova\inst{1} \and
Wiro Niessen\inst{1,5} \and
Marleen de Bruijne\inst{1,6}}

\institute{Department of Radiology \& Nuclear Medicine, Erasmus University Medical Center, Rotterdam, the Netherlands \and
Ecole des Mines de Saint-Etienne, Saint-Etienne, France \and
Ecole Centrale Marseille, Marseille, France \and
Department of Epidemiology, Erasmus MC, Rotterdam, the Netherlands \and
Department of Imaging Physics, Faculty of Applied Science, TU Delft, Netherlands \and
Department of Computer Science, University of Copenhagen, Denmark}

\maketitle

\begin{abstract}
Scoliosis is a condition defined by an abnormal spinal curvature. For diagnosis and treatment planning of scoliosis, spinal curvature can be estimated using Cobb angles. We propose an automated method for the estimation of Cobb angles from X-ray scans. First, the centerline of the spine was segmented using a cascade of two convolutional neural networks. After smoothing the centerline, Cobb angles were automatically estimated using the derivative of the centerline. We evaluated the results using the mean absolute error and the average symmetric mean absolute percentage error between the manual assessment by experts and the automated predictions. For optimization, we used 609 X-ray scans from the London Health Sciences Center, and for evaluation, we participated in the international challenge "Accurate Automated Spinal Curvature Estimation, MICCAI 2019" (100 scans). On the challenge's test set, we obtained an average symmetric mean absolute percentage error of 22.96.
\end{abstract}

\section{Introduction}

Diagnosis and treatment planning of Adolescent idiopathic scoliosis (AIS) relies on the estimation of the spinal curvature, which can be measured using Cobb angles. We propose an automated method to measure Cobb angles in X-ray scans. In our approach, the centerline of the spine was automatically segmented using cascaded neural networks, that were optimized end-to-end, i.e. trained simultaneously. While the first network focused on the segmentation of the spine, the second network focused on the extraction of the centerline of the spine, using the results of the first network. The centerline was then smoothed and its derivative was computed to obtain tangents and estimate the Cobb angles.

Unlike our approach, most automated methods for estimating Cobb angles rely on the segmentation of the individual vertebrae or the detection of their corners \cite{Wu2017,Horng2019,Mukherjee2014}. Recently, Wu et al. \cite{Wu2017} predicted the position of landmarks indicating the corners of vertebrae using convolution neural networks (CNNs). The authors proposed BoostNet, a layer architecture to reduce intra-class variance of feature embeddings. They also designed a structured output for their networks, which incorporates respective positions of landmarks in a connectivity matrix. Wu et al. \cite{Wu2017} achieved very accurate results in the automated localization of vertebrae landmarks, but did not measure Cobb angles in their paper.
Horng et al. \cite{Horng2019} recently proposed to measure the spinal curvature by first isolating spine region and detecting each vertebra using image processing, subsequently segmenting vertebrae with a U-net like network \cite{Ronneberger2015}, and measuring Cobb angles using the vertebrae segmentations.
Similarly to our approach, Okashi et al. \cite{Okashi2017} measured Cobb angles directly from the centerline of the spine. In their work, the centerline was extracted using a complex hand-engineered image processing algorithm.

\section{Methods}
\label{sec:methods}

\begin{figure}[!b]
\centering
\includegraphics[height=2.3cm]{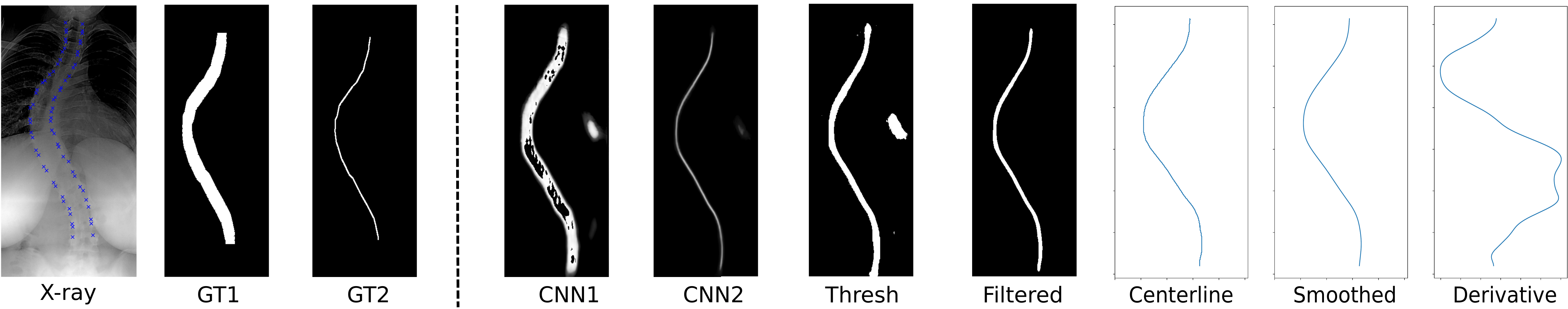}
\caption{\textbf{Spine segmentation and curvature computation.} Top row, left to right: an input X-ray scan with corners of the vertebrae manually indicated in blue (\textit{X-ray}); spine segmentation by connecting vertebral corners (\textit{GT1}) and centerline (\textit{GT2}), used for training the cascaded networks. \textit{CNN1} and \textit{CNN2} are the outputs of the cascaded networks: the spine and spine centerline segmentations, respectively. The bottom row illustrates the post-processing pipeline: thresholding the centerline segmentation (\textit{Thresh}); removing small connected components (\textit{Filtered}); extracting the spine centerline curve (\textit{Centerline}); centerline smoothing using heat equation (\textit{Smoothed}); computing the derivative of the centerline (\textit{Derivative}), which was subsequently used to compute Cobb angles.}
\label{fig:principle}
\end{figure} %

\subsection{Datasets}
\label{sec:dataset}

The method was optimized with a publicly available dataset of 609 spinal anterior-posterior X-ray images \cite{Wu2017} with 17 vertebrae of the thoracic and lumbar regions manually indicated using landmarks at the four corners thereof. Cobb angles were computed using the landmarks. To evaluate our method, we competed in the Accurate Automated Spinal Curvature Estimation (AASCE) challenge \footnote{https://aasce19.grand-challenge.org/Home/} hosted by the 22nd International Conference on Medical Image Computing and Computer Assisted Intervention. The test dataset contained 100 X-ray scans. These images were different from the training images, which focused on the spine, in that they displayed areas of the neck and shoulders. We manually cropped the images of the evaluation dataset so that they show the same region as images of the training dataset. In addition, this dataset was substantially different from the training dataset in that the majority of its images had small Cobb angles (low curvature), while Cobb angles in the training dataset were more uniformly distributed. As additional preprocessing, we normalized the intensities of all images by dividing by the maximum intensity value of each image individually.

As ground truths, the challenge organisers provided three Cobb angles that were manually measured: the \textit{major Cobb angle} which estimates the highest overall spinal curvature (the Cobb angle usually reported in the litterature); the \textit{upper Cobb angle}, which is the Cobb angle measuring the highest spinal curvature in the spine region above the region of highest overall curvature; and the \textit{lower Cobb angle}, which is the Cobb angle in the spine region under the region of highest curvature.

\begin{figure}[!b]
\centering
\includegraphics[height=1.9cm]{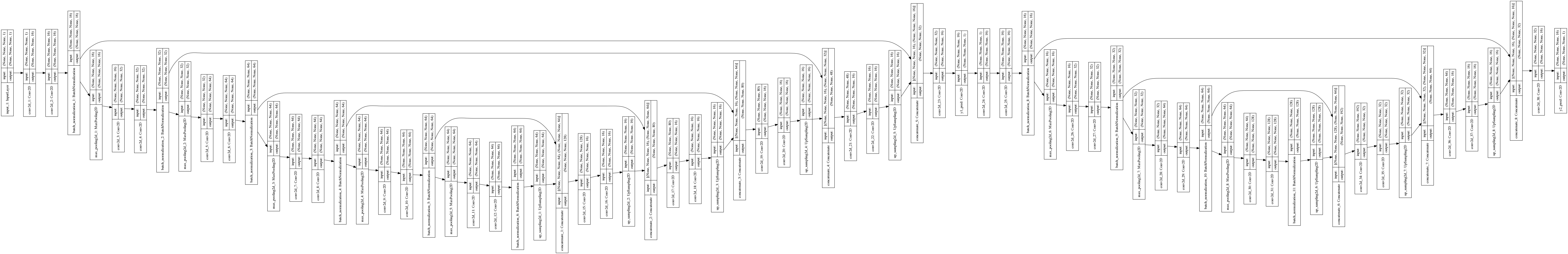}
\caption{\textbf{Cascaded Networks Architecture.} The last number in each layer is the number of feature maps.}
\label{fig:arch}
\end{figure} %

\subsection{Centerline extraction with neural networks}
Two cascaded convolutional neural networks were used to extract the centerline of the spine. Their architecture was that of a U-Net \cite{Ronneberger2015} with fewer feature maps and batch normalization layers \cite{Ioffe2015} before each pooling layer (Figure \ref{fig:arch}). The networks were optimized end-to-end, simultaneously. The first network was given the X-ray scan as input and was optimized to segment the complete spine, while the second network was given the output of the first network as input and was optimized to segment only the centerline. In earlier experiments, we used a single network to segment the centerline directly, which led to parts of the centerline -- often areas with low contrast -- not being segmented. Ground truth segmentations of the complete spine and the centerline were automatically computed using the landmarks provided in the training dataset.

The networks were optimized using Adadelta optimizer \cite{Zeiler2012} and mean squared error loss function over voxel intensities between the binary ground truth segmentations of the spine and centerline and the segmentations predicted by the network. Images were augmented online during training, with random rotation, translation, and horizontal flipping, addition of Gaussian noise, and varying brightness and contrast.

\subsection{Postprocessing}
\paragraph{Curve extraction.}

The network output was binarized using a low threshold (0.25) to ensure continuity of the centerline, followed by small (smaller than 40x140 pixels) connected components removal. Borders of the centerline were detected to compute two curves. Points in the middle of both curves were then selected to model the centerline curve. In our experiments on the validation set, this was the most robust approach to extract the curve in the training dataset. 

\paragraph{Smoothing with heat equation.} Centerline curves extracted from the images showed some local noise due to image resolution and errors in the segmentation. To compute more accurate derivatives, the centerline was smoothed using the heat equation, solved with Euler method. There were two parameters: the heat transfer coefficient, which was set to $0.01$ and the number of iterations in Euler's method set to $65000$, which had to be large enough and was tuned on the training set. We did not experiment with varying the heat transfer coefficient.

\paragraph{Computation of Cobb angles.} Similarly to Horng et al. \cite{Horng2019}, we computed Cobb angles as  
\begin{equation}
\phi = \frac{180}{\pi} \left| arctan \left( \frac{T(p_{R,M}) - T(p_{R,m})}{1+T(p_{R,M}) \cdot T(p_{R,m})} \right)\right|,
\end{equation}
where $T(p)$ is the tangent slope of the centerline at point $p$, $p_{R,M}$ is the point with the maximum slope in region $R$, and $p_{R,m}$ is the point with the minimum slope in $R$. We computed the derivative of the centerline to obtain the tangent of the centerline in every point. Then, we selected 19 points evenly spaced over the derivative of the centerline (“Derivative” in Figure \ref{fig:principle}) as an approximation of the location of the interstices between the 17 annotated vertebrae, and considered values of the slope of the tangent, $T(p)$, only in those points. Considering tangents over the entire curve would quantify its curvature more accurately but would be less comparable to the manually measured Cobb angles.

To compute the major angle, the entire centerline was used as the region of interest $R$. The upper and lower angles were computed in the regions above and below the region of the highest curvature, defined as the region between $p_{R,m}$ and $p_{R,M}$.

\begin{table}[!b]
\centering
\caption{\textbf{Automated predictions of Cobb angles.} \textit{Crop} is whether vertebrae at the top and bottom of the image were removed from the image, \textit{Nbr Models} is the number of averaged models, and \textit{Augmentation} is the type of data augmentation on image intensities, where \textit{B} \& {C} are changes of brightness and contrast over the whole image. \textit{MAE 1, 2, 3} are the mean absolute error for the major, upper, and lower Cobb angle in degrees, \textit{MAE Avg} is the MAE averaged over the three angles, \textit{SMAPE} is defined in Equation (\ref{eq:smape}) and computed over the left out test set of the public dataset (Section \ref{sec:dataset}), \textit{SMAPE Lb} was computed on the evaluation dataset and reported on the challenge's leaderboard.}\label{table:results}
\resizebox{\textwidth}{!}{%
\begin{tabular}{ccc|cccccc}
Crop & Nbr Models & Augmentation & MAE 1 & MAE 2 & MAE 3 & MAE Avg & SMAPE & SMAPE Lb \\ \hline
No   & 2          & Gaussian     & 5.28  & 5.39  & 6.91  & 5.86    & 26.45 & 27.78    \\
No   & 4          & Gaussian     & 4.95  & 5.13  & 6.55  & 5.54    & 24.27 & 25.79    \\
Yes  & 4     & B \& C       & 3.43  & 3.91 & 5.39  & 4.24    & 21.03 & 23.94   \\
Yes & 4 \& 7    & B \& C       & 3.22  & 3.75 & 5.23  & 4.07    & 20.55 & 23.67 \\
Yes & 4 \& 7 \& 4 \& 4 \& 5   & B \& C       & 3.18  & 3.6 & 5.19  & 3.99    & 20.4 & 22.96
\end{tabular}
}
\label{table:results}
\end{table}

\subsection{Ensemble of ensembles}
To improve the generalization performance of our models, we created ensembles of prediction models optimized on different subsets of the training set. The dataset of 609 X-ray scans was randomly split in 10 subsets of 61 scans each. At the beginning, one of these subsets was selected as testing set and left out of the optimization process. Models were then optimized using different combinations of the nine remaining subsets, and their results were averaged.
Models' predictions were averaged at two stages of the pipeline: centerline segmentation maps of several models were averaged to compute single Cobb angles, and Cobb angles predicted by several ensembles were averaged too, acting as an ensemble of ensembles. Results with varying size of ensembles are reported in Table \ref{table:results}. Ensembles with the best results on the left out set of 61 scans were then used for submission on the challenge test set.

\section{Experiments}

We evaluated the performance of our model by competing in the AASCE challenge (Section \ref{sec:dataset}). The AASCE challenge uses the symmetric mean absolute percentage error (SMAPE) averaged over the three Cobb angles as main metric. For all three Cobb angles, the SMAPE was computed as:
\begin{equation}
\label{eq:smape}
SMAPE = \frac{200}{n} \sum_{i=1}^{n}\frac{\sum_{j=1}^{3}|A_{i,j}-B_{i,j}|}{\sum_{j=1}^{3} A_{i,j} + B_{i,j}},
\end{equation}
where $n$ is the number of X-ray scans in the set, $A_{i,j}$ is the manual assessment of the Cobb angle $j$ for the scan $i$, and $B_{i,j}$ the Cobb angle predicted by the automated method.

To get more insights we also evaluated our results on a subset of the public dataset used for training (Section \ref{sec:dataset}) with additional metrics. This second test set contained 61 X-ray scans randomly drawn at the beginning of the experiments and not used for optimization. On this set, in addition to the SMAPE, we also computed mean absolute error (MAE) between the manual and automated assessments of Cobb angles. All results are shown in Table \ref{table:results}.

\section{Discussion and Conclusion}
We proposed an automated method for estimation of Cobb angles from X-ray scans. Contrary to most other approaches, our approach measured Cobb angle directly from the centerline of the spine without requiring the segmentation of individual vertebrae \cite{Wu2017,Horng2019,Mukherjee2014}.

Our method has several limitations. In the training dataset all images were initially cropped relatively close to the spine, while in the evaluation dataset the initial regions of interest was substantially larger. This difference caused major issues for the automated spine segmentation, and required manual cropping of the region of interest. Our method also has a long running time due to the smoothing with Euler method. A similar effect may be achieved using Gaussian smoothing. Finally, Cobb angles are not a linear measurement of severity of AIS. The severity actually increases exponentially with the Cobb angle. This is not reflected in our evaluation criteria. Using Cobb angles to assess AIS severity is also in itself limited, as Cobb angles are measured in 2D while AIS is a three-dimensional condition \cite{Giannoglou2016}. Using 3D measurements instead could help to assess AIS severity more accurately.

Intrarater or interrater variability have not been measured in our dataset. By inspecting results reported by Horng et al. \cite{Horng2019}, we computed that, on their set of 32 X-ray scans, the intrarater variability in the manual measurement of Cobb angle by an expert observer had a MAE of 2.0 degrees and interrater variability had a MAE of 3.87 degrees on average. Our results are in-between these intrarater and interrater variabilities, with a MAE of 3.18 degrees and Pearson correlation coefficient of 0.95 between manual and automated measurements of the major Cobb angle in our dataset. Consequently, our method might be suited to replace manual assessment of Cobb angles in clinical practice.

\section{Acknowledgments}
This research was funded by the Netherlands Organisation for Health Research and Development (ZonMw) Project 104003005, with additional support of Netherlands Organisation for Scientific Research (NWO) project NWO-TTW Perspectief Programme P15-26.

\end{document}